\documentclass[a4paper]{spie}
\usepackage{graphicx}
\usepackage{amssymb}
\usepackage{hyperref}

\title{The GRAVITY instrument software / High-level software}

\author{Leonard Burtscher\supit{a}\footnote{\,\,\,burtscher@mpe.mpg.de}, Ekkehard Wieprecht\supit{a}, Thomas Ott\supit{a}, Yitping Kok\supit{a}, \c{S}enol Yaz{\i}c{\i}\supit{b}, Narsireddy Anugu\supit{c}, Roderick Dembet\supit{d}, Pierre F\'edou\supit{d}, Sylvestre Lacour\supit{d}, J\"urgen Ott\supit{e}, Thibaut Paumard\supit{d}, Vincent Lapeyrere\supit{d}, Pierre Kervella\supit{d}, Roberto Abuter\supit{f}, Eszter Pozna\supit{f}, Frank Eisenhauer\supit{a}, Nicolas Blind\supit{a}, Reinhard Genzel\supit{a}, Stefan Gillessen\supit{a}, Oliver Hans\supit{a}, Marcus Haug\supit{a}, Frank Hau\ss mann\supit{a}, Stefan Kellner\supit{a}, Magdalena Lippa\supit{a}, Oliver Pfuhl\supit{a}, Eckhard Sturm\supit{a}, Johannes Weber\supit{a}, Antonio Amorim\supit{c}, Wolfgang Brandner\supit{g}, Karine Rousselet-Perraut\supit{h}, Guy S. Perrin\supit{d}, Christian Straubmeier\supit{b}, Markus Sch\"oller\supit{f}, 
\skiplinehalf
\supit{a} Max-Planck-Institut f\"ur extraterrestrische Physik, Postfach 1312, Gie\ss enbachstr., 85748 Garching, Germany\\
\supit{b} 1. Physikalisches Institut, Universit\"at K\"oln, Z\"ulpicher Str. 77, 50937 K\"oln, Germany\\
\supit{c} Universidade do Porto - Faculdade de Engenharia, Departamento de Engenharia F\'isica, Laborat\'orio SIM, Rua Dr. Roberto Frias, s/n, 4200-465 Porto, Portugal\\
\supit{d} Observatoire de Paris / LESIA, 5, place Jules Janssen, 92195 Meudon Cedex, France\\
\supit{e} redlogix Software \& System Engineering GmbH, Talhofstr. 32a, 82205 Gilching, Germany\\
\supit{f} European Southern Observatory, Karl-Schwarzschild-Str. 2, 85748 Garching, Germany\\
\supit{g} Max-Planck-Institut f\"ur Astronomie, K\"onigstuhl 17, 69117 Heidelberg, Germany\\
\supit{h} IPAG, 414 rue de la piscine, Domaine universitaire, 38400 Saint Martin d'H\`{e}res, France}

\begin{document}
\maketitle

\begin{abstract}
GRAVITY is the four-beam, near- infrared, AO-assisted, fringe tracking, astrometric and imaging instrument for the Very Large Telescope Interferometer (VLTI). It is requiring the development of one of the most complex instrument software systems ever built for an ESO instrument. Apart
from its many interfaces and interdependencies, one of the most challenging aspects is the overall performance and stability of this complex system. The three infrared detectors and the fast reflective memory network (RMN) recorder
contribute a total data rate of up to 20 MiB/s accumulating to a maximum of 250 GiB of data per night. The detectors, the two instrument Local Control Units (LCUs) as well as the five LCUs running applications under TAC (Tools for Advanced Control)
architecture, are interconnected with fast Ethernet, RMN fibers and dedicated fiber connections as well as signals for the time synchronization. Here we give a simplified overview of all subsystems of GRAVITY and their interfaces and discuss two examples of high-level applications during observations: the acquisition procedure and the gathering and merging of data to the final FITS file.

\end{abstract}


\section{Introduction}
GRAVITY\cite{eisenhauer2014} is an interferometer under development for the Very Large Telescope Interferometer (VLTI) that will combine the light of four telescopes in the near-infrared, resulting in images with a resolution of about 2 milli-arcseconds. Additionally it is designed to provide astrometry with a precision of 10 micro-arcseconds. Its main science goal is to detect motions close to the event horizon of the Galactic Center, but it is also well suited to study the kinematics of the Broad Line Region of Active Galactic Nuclei, disks around young stellar objects, evolved stars and even asteroids in our solar system. The entire observing sequence is controlled by the INstrument Software (INS) which is presented in two papers in these proceedings. Paper I focuses on the hardware aspects\cite{ott2014}; here we give an overview of the software implementation with an emphasis on the high-level aspects.

\section{Instrument software: Overview}
\label{sec:ins}

The instrument software of GRAVITY is a distributed application in the framework of the ESO VLT software\footnote{see http://www.eso.org/projects/vlt/sw-dev/wwwdoc/VLT2011/dockitVLTCore.html for technical details on ESO technologies mentioned in this article} that runs on a number of Linux workstations and real-time computers. Most computers run one or more ``environments''. These are dedicated applications in charge of specific control tasks, e.g. the ``workstation environment'' (on the instrument workstation) includes processes for observing control while the different ``ICS environments'' (on the LCUs close to the hardware) most importantly run a server process to forward device commands to the proper hardware. The environments also run a number of helper processes and a messaging system to command the various control processes; the state of devices (e.g. shutter closed) as well as transient data (e.g. a pixel to visibility matrix, P2VM) are stored in an online database whose contents can be ``scanned'' between environments to share the data.

On the highest level and during observations, the instrument is controlled through observing templates which command the GRAVITY Observation Software (OS), a server process that runs on the instrument workstation and controls its subsystems: the Instrument Control Software (ICS) and the Detector Control Software (DCS). It also controls real-time applications running under Tools for Advanced Control (TAC) and the Interferometer Supervisory System (ISS) which distributes these commands to all telescopes and the VLTI infrastructure.

\subsection{The Observation Software (OS) and the VLTI interface}
\label{sec:ins:os}

\begin{figure}
	\begin{center}
		\begin{tabular}{c}
			\includegraphics[width=10cm]{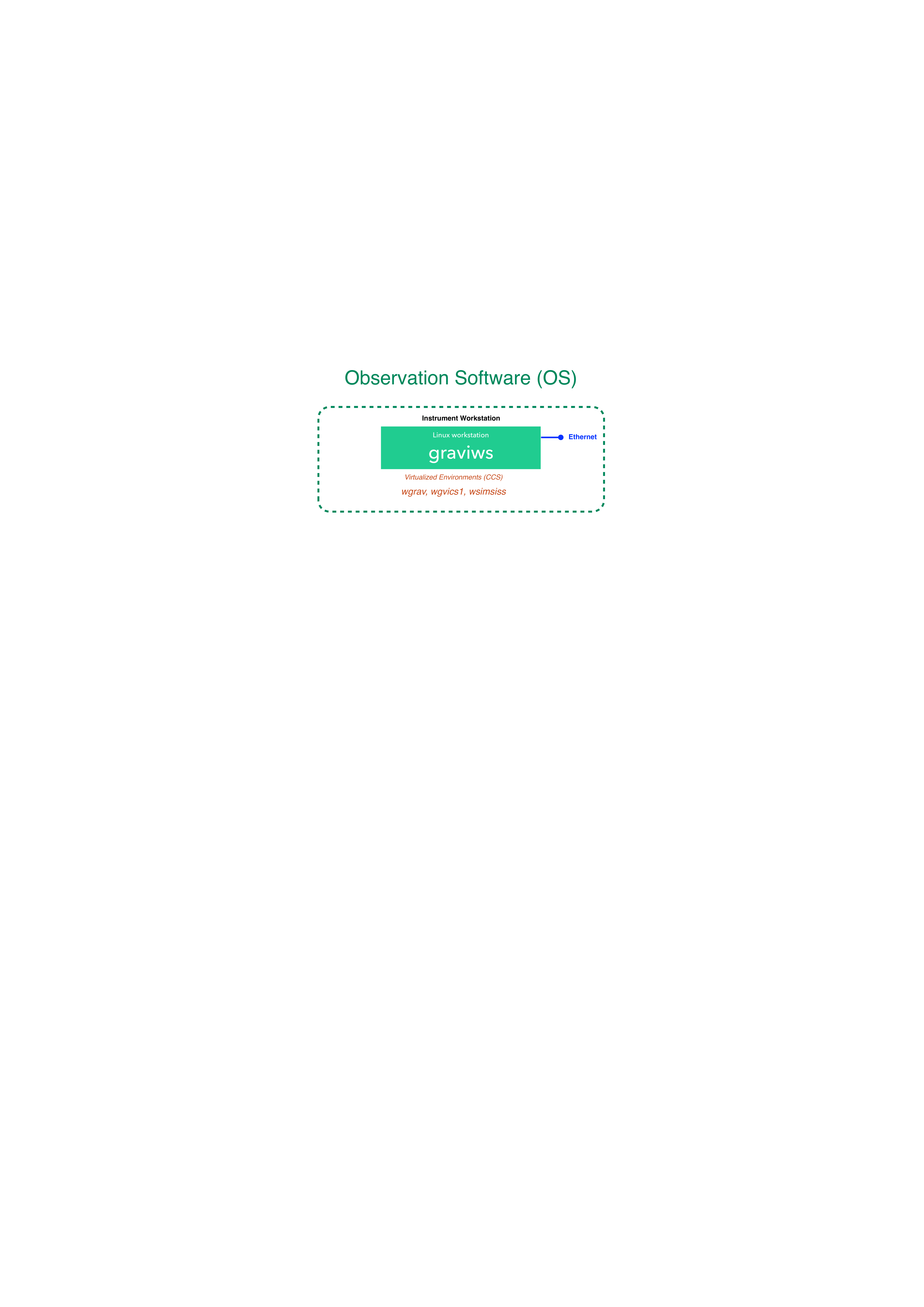}
		\end{tabular}
	\end{center}
	\caption{\label{fig:ins:os} The Observation Software consists of processes running on the instrument workstation. It is connected to the instrument network using Ethernet.}
\end{figure}

\begin{figure}
	\begin{center}
		\begin{tabular}{c}
			\includegraphics[width=10cm]{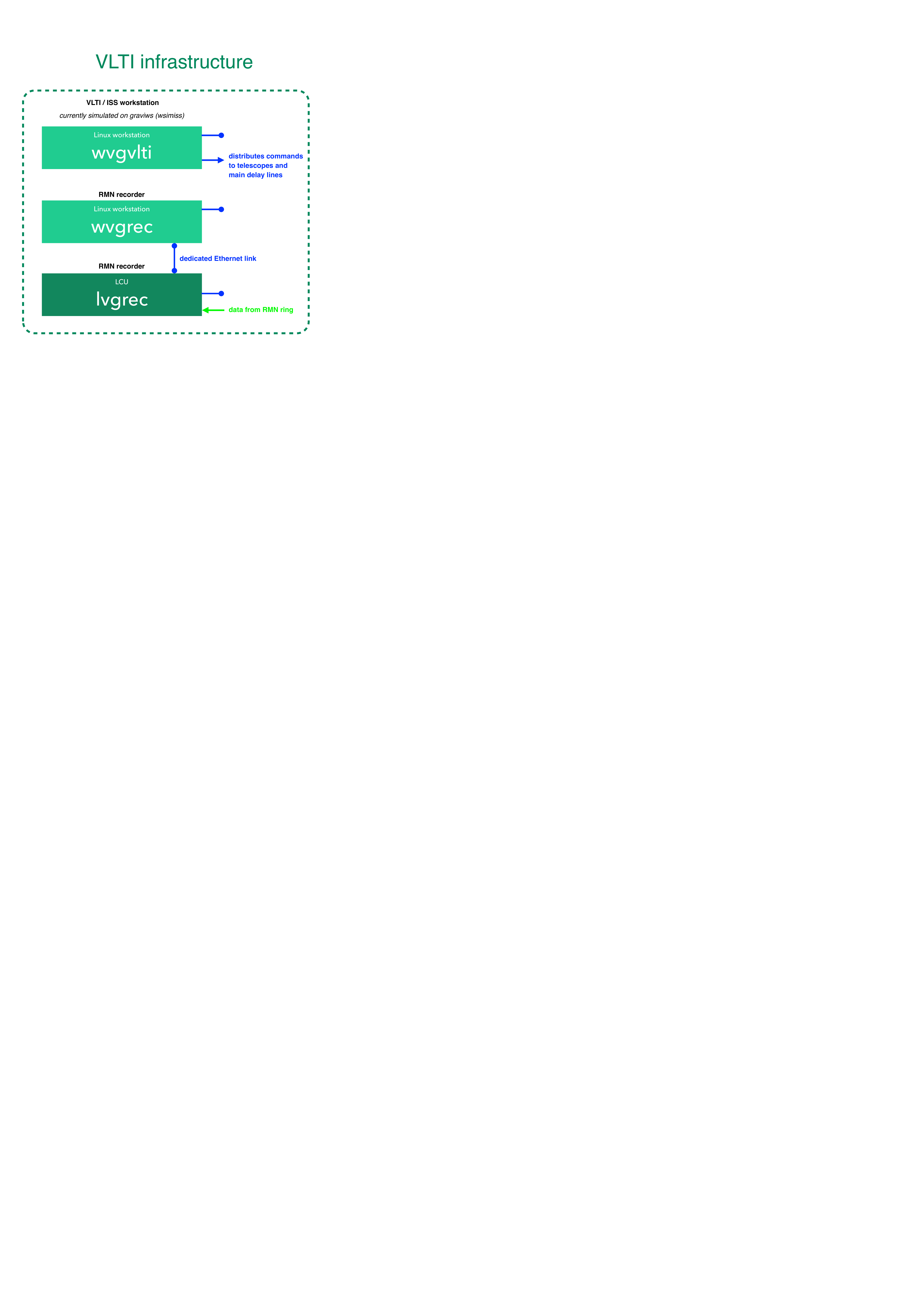}
		\end{tabular}
	\end{center}
	\caption{\label{fig:ins:vlti} The Interferometer Supervisory System (ISS) provides the interface between the instrument and the VLTI. Except for the RMN recorder LCU and workstation, all parts of the VLTI are currently simulated on the instrument workstation. Blue arrows symbolize Ethernet connections; green arrows symbolize RMN network connection. To ensure fastest data transfer, there is a dedicated direct Ethernet link between the RMN recorder LCU and workstation.}
\end{figure}

The Observation Software (Fig.~\ref{fig:ins:os}) and other high-level tasks are implemented as server processes and run physically on the ``instrument workstation'', a standard Linux server that is connected via Ethernet to the instrument network. The machine currently runs three ``environments'':

\begin{itemize}
	\item {\bf wgrav} \quad Workstation environment containing all of the processes mentioned below
	\item {\bf wgvics1} \quad Environment controlling those Instrument Control Software (ICS) devices that are directly connected to the network via Ethernet (such as some temperature controllers) and do not require a dedicated Local Control Unit (LCU).
	\item {\bf wsimiss} \quad Simulation for the actual VLTI environment (wvgvlti), see Fig.~\ref{fig:ins:vlti}
\end{itemize}

The main processes that are running on the instrument workstation and have been developed specifically for use in GRAVITY are:
\begin{itemize}
	\item {\bf gvacqControl} \quad Acquisition Camera sensor -- calculates e.g. the position of stars in the field\cite{anugu2014}.
	\item {\bf gvttpControl} \quad Acquisition Camera control loop -- actual control loop, taking e.g. star positions from {\rm gvacqControl}, computes offset values and sends them as setpoints to the tip / tilt and piston control process running on {\rm lgvttp} (see below)
	\item {\bf gvdlControl} \quad responsible for offloading piston from internal piezos to the internal delay lines in the calibration unit -- in use when running the system in calibration or testing mode
	\item {\bf gvspcControl} \quad sensor and control loop process for group delay guiding of the science fringes
	\item {\bf gvctuControl} \quad coordinate transform unit -- server process that is used to transform the various coordinate systems in use in GRAVITY (e.g. sky [arc seconds], Tip Tilt Piston mirror piezos [Volts], Position Sensitive Diodes [Volts], Roof Prism [microns], Acquisition Camera [pixels], Fiber Positioners A /B (for science and fringe tracker object) [Volts]). Some transformations are trivial (e.g. acq camera to sky is just the pixel scale), others are well described by a simple 2x2 rotation matrix (e.g. tip tilt piston to acq camera). Not all conversions that are possible are defined as direct conversions. E.g. to convert from Fiber Positionier voltage to arcseconds on sky, we calibrate the conversion from Fiber Positioner voltages to pixels on acquisition camera. With the known pixel scale this can then easily be converted to an angle on-sky. The server process knows about the conversion sequence and is configured such that it directly gives the answer to any directly or indirectly defined conversion.
\end{itemize}

Apart from these processes that have been created specifically for GRAVITY, there is also a large number of processes that belong to the standard VLT software environment. These processes have been adapted (e.g. by overloading functions that are empty by default) for our use. Some of the relevant high-level processes are:
\begin{itemize}
	\item {\bf gviControl} \quad control process for Instrument Control Software (ICS)
	\item {\bf gvoControl} \quad control process for Observation Software (OS), distributes commands to subsystems. The base version of this process is supplied by ESO with the Base Observation Software Stubb (BOSS) and its adaptation for VLTI (bossVLTI)\cite{pozna2008,pozna2012,pozna2014} and has been altered by us to handle the instrument-specific FITS files (see below).
	\item {\bf bob} \quad The Broker for Observing Blocks handles so called Observing Blocks (OBs). These are pre-defined observation sequences consisting of one or more ``Templates'' (Tcl/Tk scripts + parameter files, so called Template Setup Files (TSF)). The templates in turn can have user-definable parameters which are edited by the observer to setup the observation (e.g. coordinates of the object and instrument setup). Each ``OB'' consists of one or more templates.
	\item {bossArchiver\_gvo} \quad instance of the bossArchiver for GRAVITY; merges images, tables and headers to final FITS file. The merge process is configured using so called Archiver Reference Files (ARF), that specify what files are to be merged and deleted. The ARF file is produced by the observation control process (gvoControl) at the end of the exposure and the archiver is then called to handle the arf file and merge the instrument data to the final product.
\end{itemize}

\subsection{The Instrument Control Software (ICS) and real-time applications using TAC}
\label{sec:ins:ics}

\begin{figure}
	\begin{center}
		\begin{tabular}{c}
			\includegraphics[width=10cm]{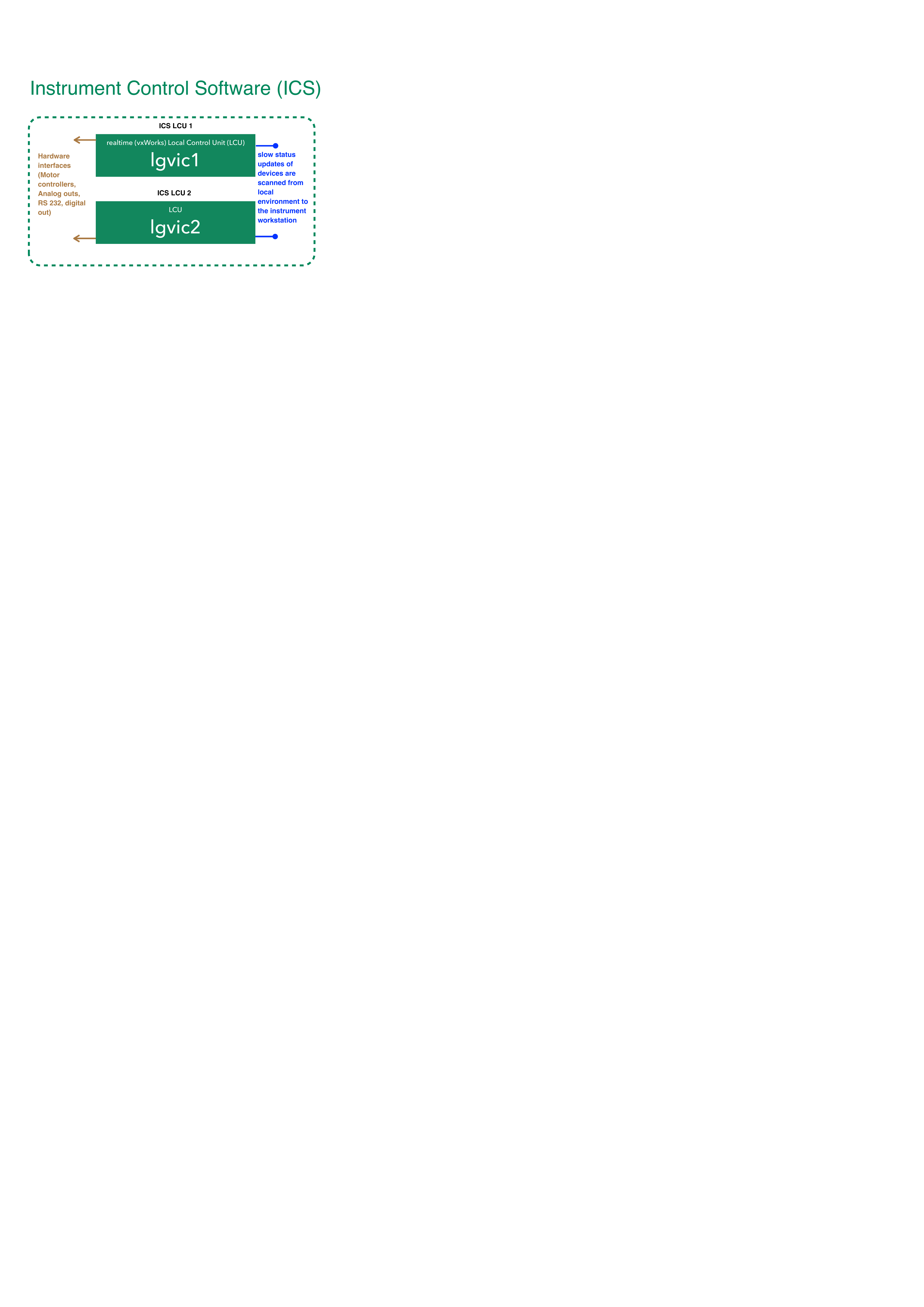}
		\end{tabular}
	\end{center}
	\caption{\label{fig:ins:ics} The Instrument Control Software. While the control server runs on the instrument workstation, the actual devices are commanded through dedicated LCUs. Blue arrows symbolize Ethernet connections; brown arrows symbolize other interfaces.}
\end{figure}

\begin{figure}
	\begin{center}
		\begin{tabular}{c}
			\includegraphics[width=10cm]{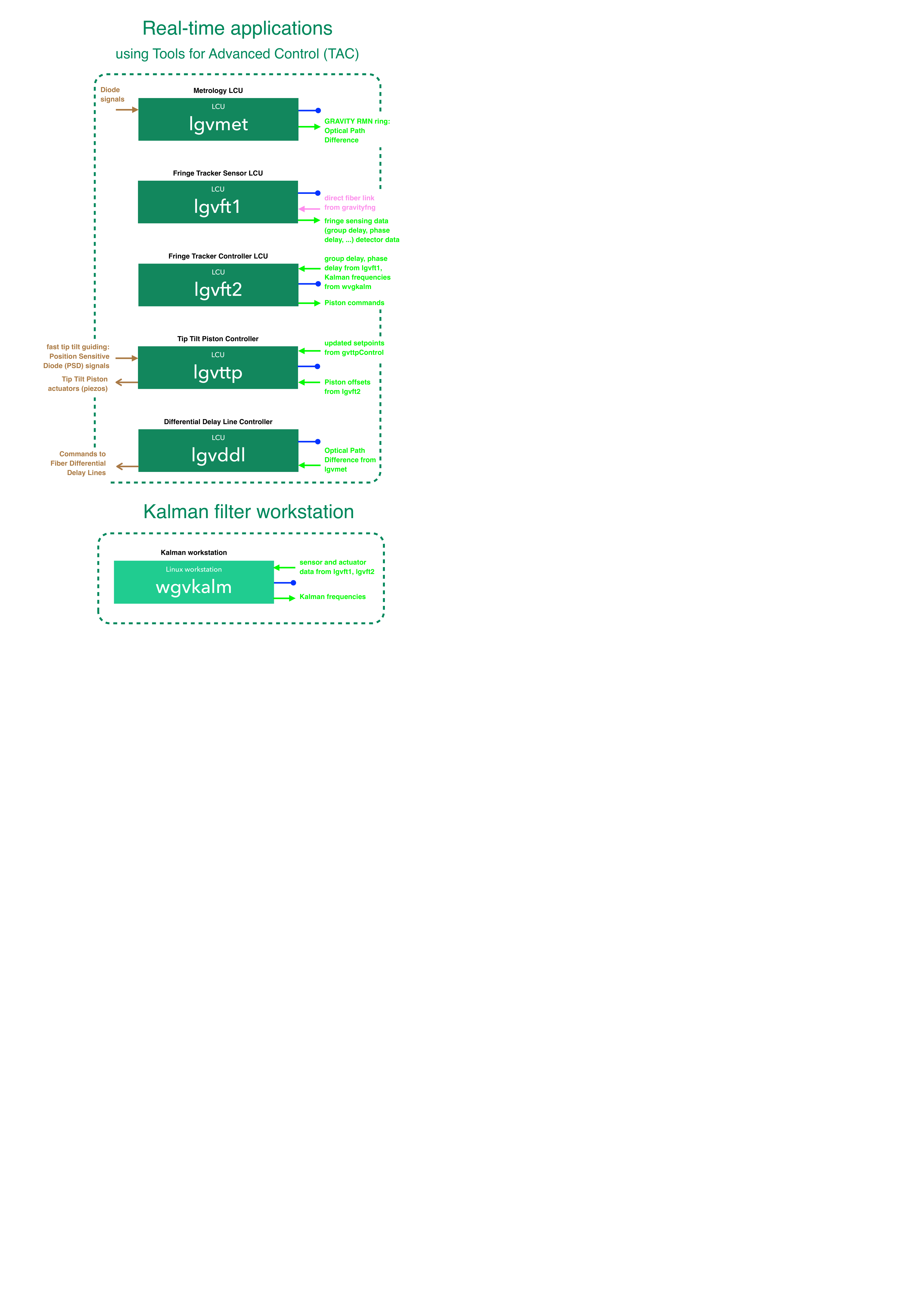}
		\end{tabular}
	\end{center}
	\caption{\label{fig:ins:tac} The GRAVITY TAC applications and communication between the real-time processes. Blue arrows symbolize Ethernet connections; green arrows symbolize RMN network connection, brown and purple arrows symbolize other interfaces.}
\end{figure}

The Instrument Control Software consists of a server process running on the instrument workstation which evaluates commands and forwards them to the correct Local Control Unit (LCU). The LCUs run a real-time operating system (vxWorks) and provide the interface to the actual hardware (see Fig.~\ref{fig:ins:ics}). GRAVITY consists of two ICS LCUs with various devices such as motors, lamps, shutters, ... attached\cite{ott2014}. GRAVITY is further making use of five additional LCUs that run Tools for Advanced Control (TAC), a framework created for real-time control processes such as the fast control loops of GRAVITY (metrology, fringe tracker, tip/tilt/piston control and differential delay line control). The communication between these LCUs (and the additional non-realtime Kalman workstation) is displayed in Fig.~\ref{fig:ins:tac}.

Communication between these TAC LCUs is handled through a fast fiber link, the Reflective Memory Network (RMN).

\subsection{The Detector Control Software (DCS)}
\label{sec:ins:dcs}

\begin{figure}
	\begin{center}
		\begin{tabular}{c}
			\includegraphics[width=10cm]{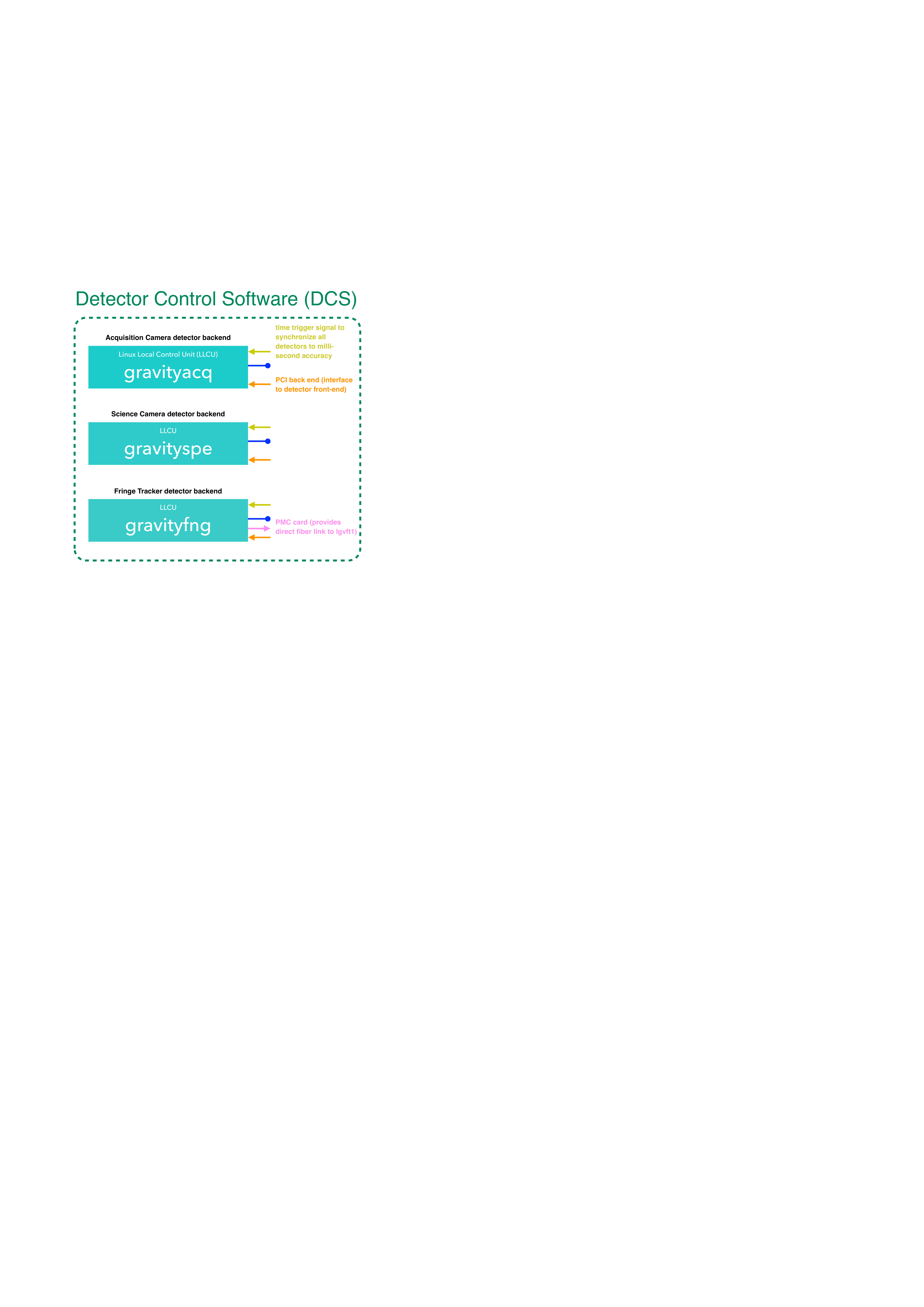}
		\end{tabular}
	\end{center}
	\caption{\label{fig:ins:dcs} The Detector Control Software. The control process runs on the instrument workstation, but the interface to the actual detector hardware is done with standard Linux PCs as local control units (so called LLCUs). Blue arrows symbolize Ethernet connections, yellow, orange and purple symbolize other interfaces.}
\end{figure}

Detector control for the three infrared detectors of GRAVITY (two HAWAII-2RG and one SELEX) uses the standard ESO NGC architecture where the data reception task runs on a Linux computer which transmits the data frame-by-frame via Ethernet to the instrument workstation where the data are displayed (in the real-time displays), evaluated (e.g. by the acquisition camera control process) and eventually saved to disk (see Fig.~\ref{fig:ins:dcs}).

The workstation connected to the SELEX detector (which is used for the fast fringe tracking) has an additional direct link to the real-time fringe tracker TAC LCUs in order to process the raw data with as little latency as possible. The data are then copied to the fast RMN network by the TAC LCU and stored to disk using a dedicated data recorder facility, the RMN recorder\cite{abuter2008} that is configured to filter and store data in the formats required by GRAVITY.

The exposures for all detectors are synchronized using a time signal provided by ESO using TIM devices (see below).

\section{A GRAVITY observation and the data product}

A GRAVITY observation will consist essentially of two templates, an acquisition and an observing template. Additionally, various calibration templates may be needed, depending on the actual science observation.

The acquisition sequence for the off-axis AO and off-axis fringe tracking case, is a staggered process that starts with the preset of the telescopes, domes and delay lines and the starting of the telescope guiding that is parallelized as much as possible (e.g. delay lines start to move at the same time as domes and telescopes start to move). After the telescope guiding has started, the star separators within the telescopes are set up such to relay one beam to the adaptive optics (AO) train and the other towards the VLTI delay lines. This includes the interactive confirmation of the two fields for all four telescopes. After the AO loop is closed and the science field is properly centered on the star separator, the fringe tracking object is selected on the GRAVITY-internal acquisition camera for all four beams. The fringe tracking star is then moved onto the fiber that feeds the fringe tracking arm of GRAVITY and the light in that arm is optimized using the internal tip-tilt mirror. Once fringes are found using the internal piston actuator (and offloading large values of piston to the main delay lines), the science object is confirmed interactively for all four beams and the light is optimized on this arm using the fiber positioners. After centering the science fringes using a group delay estimator, GRAVITY is ready for observations.

When GRAVITY is ready for observations, the observation template sets up the exposures by requesting exposure identification numbers (expoIds) for each detector. The setup also includes sending a common start time to the ``TIM'' devices. When this time is reached, these devices start sending trigger signals in the defined time-intervals to the detectors. This way, the precise time of each sub-integration can be reconstructed from the starting time, the integration time and the exposure number.

At the end of the exposure, the OS triggers the production of header files from its subsystems and writes the ``archivation reference file'' which specifies what files will be merged together to form the final FITS file with all extensions. In the standard ESO software this is done separately for each detector. For GRAVITY the requirement was, however, to merge the individual exposures and headers from all subsystems into one FITS file in order to ease the archiving process and simply to have all relevant information conveniently in a single file.

For this, we modified the observation software such that file merging is deactivated by default in order not to create partly merged files. At the end of the exposure, ``arf'' files are written depending on the current instrument configuration and the archiver process is called from within the template to merge the files. To ensure fast disk I/O, our instrument workstation is equipped with solid state drives on which the merging process is performed within a few seconds for the maximum file size of 2 GiB.

\begin{figure}
	\begin{center}
		\begin{tabular}{c}
			\includegraphics[width=10cm]{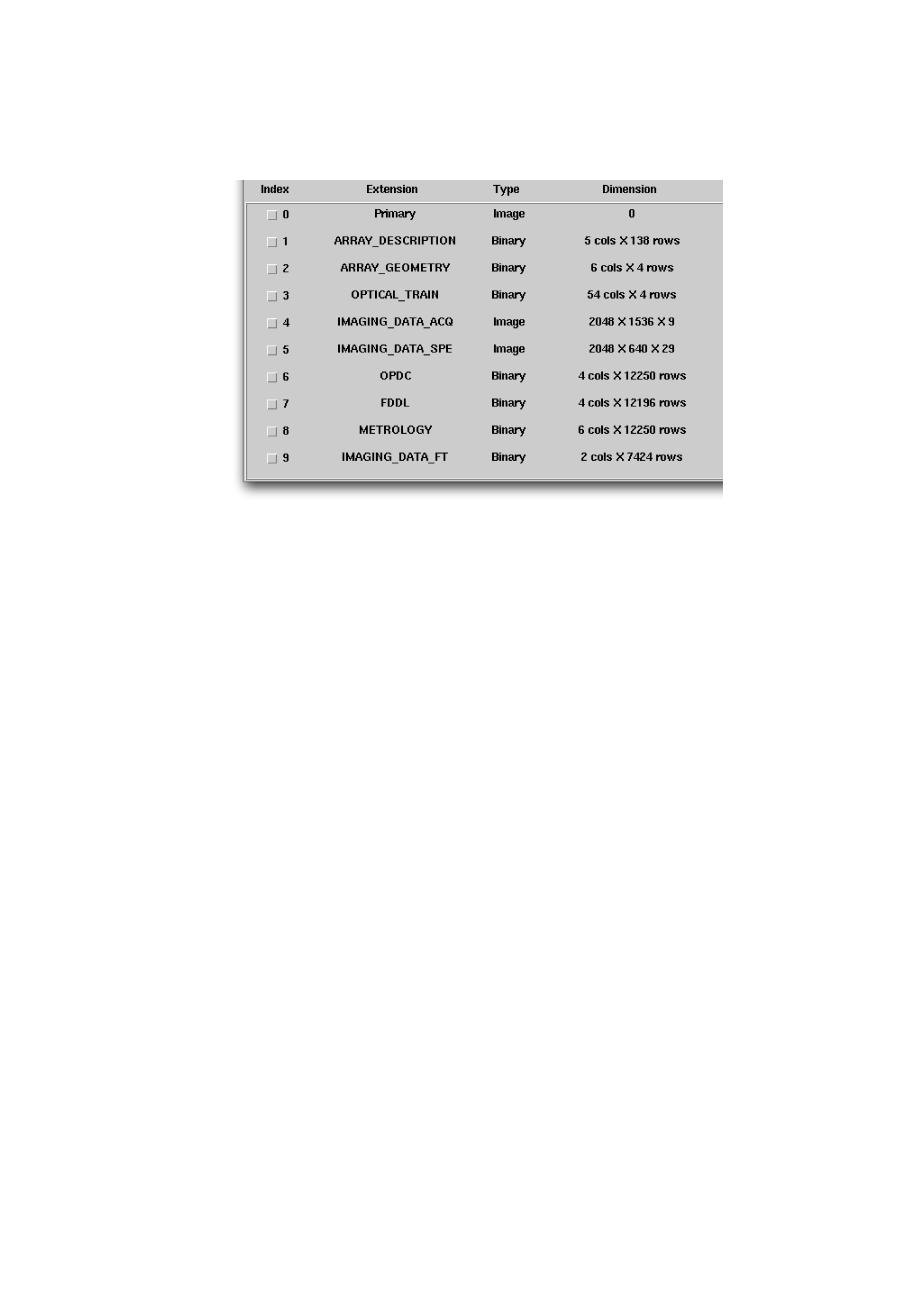}
		\end{tabular}
	\end{center}
	\caption{\label{fig:ins:fits} The GRAVITY data product: A merged FITS file with many extensions.}
\end{figure}

The merged FITS file consists of three binary tables provided by the VLTI subsystem, two image cubes (from the acquisition camera and science spectrometer detectors) as well as four binary tables recorded by the RMN recorder. Two of the RMN tables are provided by the fringe tracking application which is distributed over two LCUs, the other two are from the fiber differential delay line controller and the metrology controller (see Fig.~\ref{fig:ins:fits}).

\section{Auxiliary systems for the GRAVITY instrument development}

The GRAVITY instrument software produces technical log files containing both ``FITS logs'' (device temperatures, pressures, fan speeds, ...) as well as messages from the processes running on the various computers. Depending on the instrument and testing operations, these log files amount to a few MiB up to a few GiB per day.

\begin{figure}
	\begin{center}
		\begin{tabular}{c}
			\includegraphics[width=10cm]{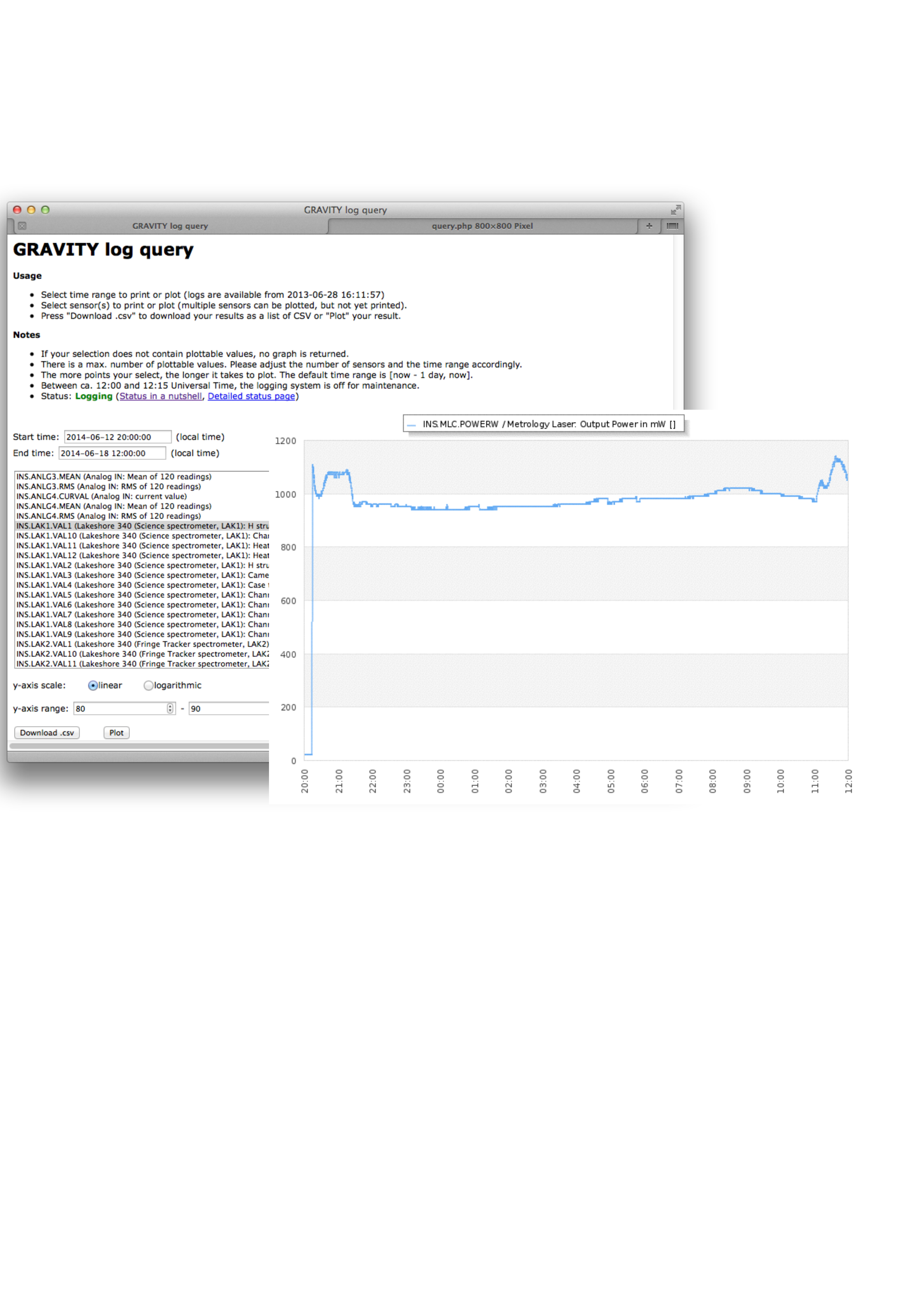}
		\end{tabular}
	\end{center}
	\caption{\label{fig:ins:glog} The GRAVITY log webpage provides easy access to the instrument logs.}
\end{figure}

We developed a log parsing and querying tool in order to easily extract and plot time sequences of relevant values, such as the output power of lasers, the temperature of some component or the amount of cooling water used. The tool consists of two parts. The back end parses the raw log file for relevant entries, formats them as SQL statements and inserts these into an SQLite database. The front end consists of a set of PHP scripts to query the database and plot the resulting values using the versatile open-source plotting library jpgraph \footnote{http://jpgraph.net}. The web interface and an example plot are shown in Fig.~\ref{fig:ins:glog}.

\bibliography{../papers}   
\bibliographystyle{spiebib}   

\end{document}